\documentclass[aps,pre]{revtex4}
\usepackage{graphicx}
\begin{document}

\title{Collective effects induced by diversity in extended systems} 

\author{Ra\'ul Toral, Claudio J. Tessone}
\affiliation{Instituto Mediterr\'aneo de Estudios Avanzados (IMEDEA), CSIC-UIB, Ed. Mateu Orfila, Campus UIB, 07122-Palma de Mallorca, Spain}
\author{Jo\~ao Viana Lopes}
\affiliation{Centro de F{\'\i}sica do Porto, Departamento de F{\'\i}sica, Faculdade de Ci\^encias, Universidade do Porto, 4169-007 Porto, Portugal}
\begin{abstract} 
We show that diversity, in the form of quenched noise, can have a constructive effect in the dynamics of extended systems. We first consider a bistable $\phi^4$ model composed by many coupled units and show that the global response to an external periodic forcing is enhanced under the presence of the right amount of diversity (measured as the dispersion in one of the parameters defining the model). As a second example, we consider a system of active-rotators and show that while they are at rest in the homogeneous case, the disorder introduced by the diversity suffices to trigger the appearance of common firings or pulses. Both effects require very simple ingredients and we expect the results presented here to be of interest in similar models.

\end{abstract}
\maketitle
%
\section{Introduction}
In dynamical systems, the elimination of some internal degrees of freedom leads to  an effective Langevin equation which contains so-called noise terms\cite{langevin}. Those terms aim to represent in a probabilistic description our lack of knowledge about the exact microscopic dynamics. Typical effects of the noise terms are the erratic movement of the Brownian particle or the disappearance of the ferromagnetic phase at high enough temperatures. Given these and other similar examples, people were led to identify noise and disorder. However, one of the most astonishing developments of the last decades in the field of stochastic processes is that noise can have a constructive role. There are now several examples of that counterintuitive role of noise: phase transitions where a more ordered phase appears when increasing the noise intensity\cite{BPT94}; stochastic resonance where the response to an external forcing improves in the presence of noise\cite{BSV:1981,NN:1981, GHJM:1998}; coherence resonance \cite{PK97} (also named as stochastic coherence \cite{stocoh}) where an optimal periodicity in the output of the dynamical system appears for the right noise value, noise sustained patterns, structures and fronts\cite{nspat,nsfro}, etc. In these and other similar examples, the constructive role appears only in a limited window range of noise intensities. Too a large noise always recovers its disordering features. Studies have been performed in simple low-dimensional systems as well as in extended systems with many degrees of freedom\cite{wio,gosan}. In the latter case, a usual assumption is that the units composing the ensemble are identical in the sense that they all have the same values for the  relevant parameters characterizing their dynamical behavior. The only difference in their detailed dynamics comes from the noise terms which take randomly different (uncorrelated) values for each of the units.

When considering extended systems \cite{gosan}, mostly in the biological sciences, the assumption of identical units is not an appropriate one, since those systems present a, sometimes large, diversity in the units. For example, cells of the same type vary in size and shape, as well as having different membrane conductivities, etc. Hence, it is not reasonable to assume that, even in the absence of noise, they will behave in the same way. Natural diversity (also known as ``variability") has been considered in several different contexts: from the quenched noise terms representing impurities in magnetic systems \cite{ma} (where the critical temperature is lowered by the presence of the impurities) to the Kuramoto model of coupled random oscillators with different natural frequencies \cite{YK:1975,YK:1984} (where synchronization is destroyed for a sufficiently large dispersion of frequencies). All these examples lead us again to identify diversity with disorder. 

However, as in the case of noise, some recent studies lead to the conclusion that diversity might have a constructive role. Some previous work has shown the existence the phase transitions \cite{BPR01}  and patterns \cite{BL03}  induced by quenched dichotomous noise. In this paper we will briefly comment on two recently found effects: resonance to an external forcing and common firing, both induced by diversity. In forced systems non-linear systems, we have proven \cite{TMTG:2006} that the global response of an extended system to an external forcing can be improved when the units of the system are not identical. In the case of global firing, we have shown \cite{TSTC:2006} that diversity can help the units of an extended system to pulse in synchrony. In both cases the basic mechanisms for the emergence of the collective behavior  (described in detail in the following sections) require only very generic properties and both effects could occur in similar systems. 

The paper is organized as follows: in section \ref{phi4} we study the $\phi^4$ model, a paradigmatic bistable model for phase transitions: in subsection \ref{phi4-dis} we show the usual role of diversity, which leads to the disappearance of the ferromagnetic order. Next, in subsection \ref{phi4-ord} we show that the global response of the model to a periodic forcing shows a maximum for a well defined value of the diversity. In section \ref{actrot} we discuss a system of active rotators, a paradigmatic model for collective excitable behavior. First, in subsection \ref{actrot-dis}, we first discuss the usual disordering role of diversity, while in subsection \ref{actrot-ord} we show that a collective behavior, in the form of synchronized pulses, appears for the right value of the diversity. The main conclusions of the paper are summarized in the final section \ref{conclusions}.

\section{Diversity-induced resonance in the $\phi^4$ model}\label{phi4}
The $\phi^4$ model (also called Ginzburg-Landau or model A) is one of the basic models in equilibrium statistical mechanics and has been used to model many physical situations, although the simplest application is to describe the paramagnetic-ferromagnetic transition that occurs as a function of the temperature. In this model, a set of real variables $x_i(t)$, $i=1,\dots,N$ are located in the sites of a regular $d$-dimensional lattice. 
\begin{equation} \label{eq:phi4}
\frac{d x_i}{dt}=ax_i-x_i^3+\frac{C}{N_i}\sum_{j\in {\cal N}_i}(x_j-x_i)+\eta_i.
\end{equation}
Here ${\cal N}_i$ denotes the set of neighboring sites with which site $i$ interacts, and $N_i$ is the number of such neighboring sites. $C$ is the coupling constant. An usual version of this model includes in ${\cal N}_i$ only the $2d$ nearest neighbors of $i$. In this paper, we will be considering the mean-field or all-to-all coupling version in which all sites interact with the same strength. Hence ${\cal N}_i$ contains all the lattice sites and $N_i=N$. The disorder $\eta_i$ is usually considered to be a white noise of intensity proportional to the temperature. The model then displays a phase transition from an ordered (ferromagnetic) phase to a disordered (paramagnetic) phase at a critical temperature $T_c$\cite{TC90}. This is the generic behavior when $a>0$ and $C>0$, the only cases considered in this paper.

As we stated in the introduction, we are interested in analyzing the role of diversity in the units $x_i$. To this end, we neglect the thermal noise. Instead, the diversity appears as quenched noise, i.e.~the values $\eta_i$ (with $i=1,\dots,N$) do not depend on time, and are independently drawn from a probability distribution $g(\eta)$. At this moment we only assume a symmetric distribution $g(\eta)=g(-\eta)$. The mean value of the distribution is $\langle \eta_i\rangle=0$ and the correlations are $\langle \eta_i\eta_j\rangle=\sigma^2\delta_{ij}$. The standard deviation $\sigma$ is a measure of the diversity.

\subsection{The disordering role of diversity}\label{phi4-dis}

\begin{figure}
\begin{center}
\includegraphics[width=7cm]{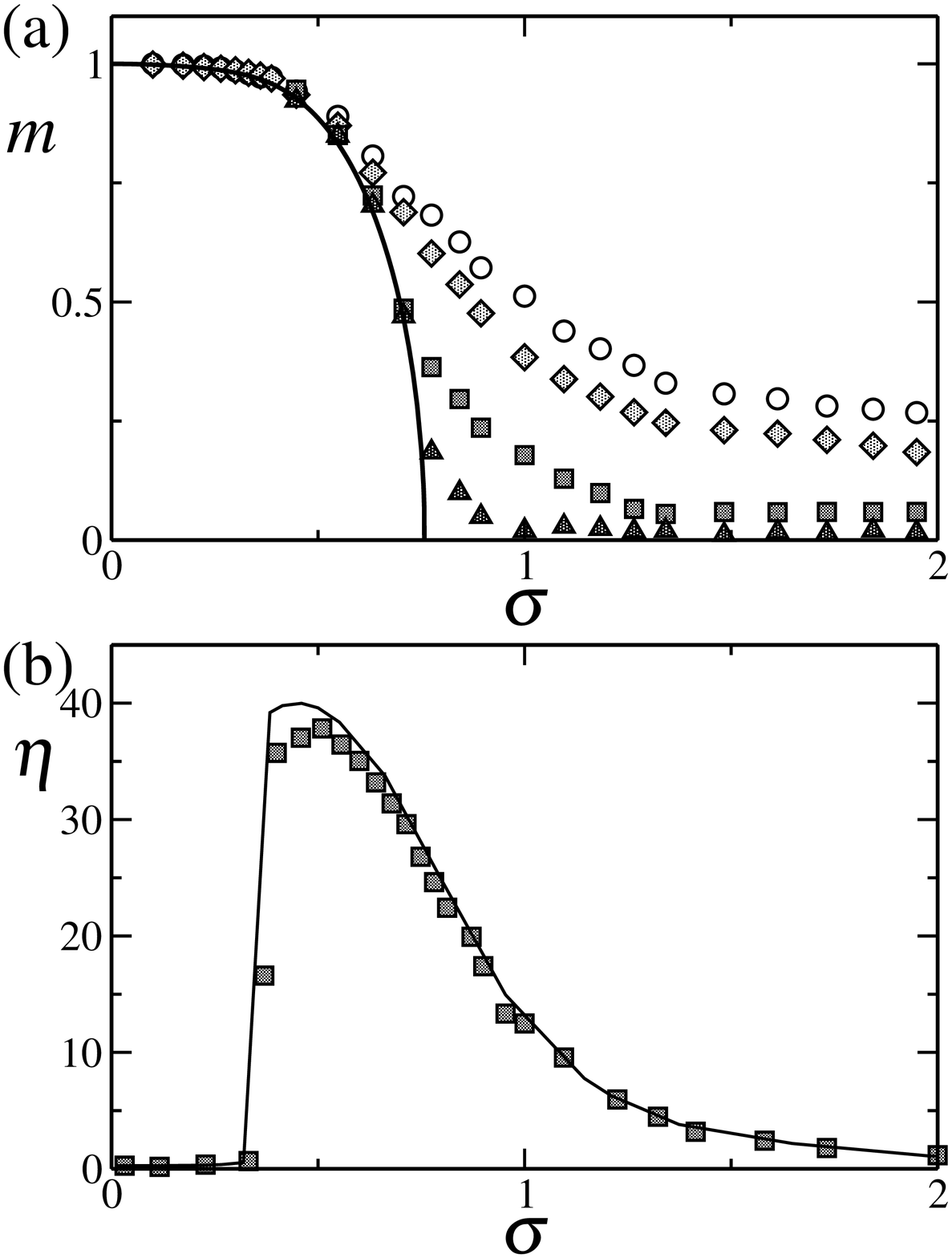}
\includegraphics[width=7cm]{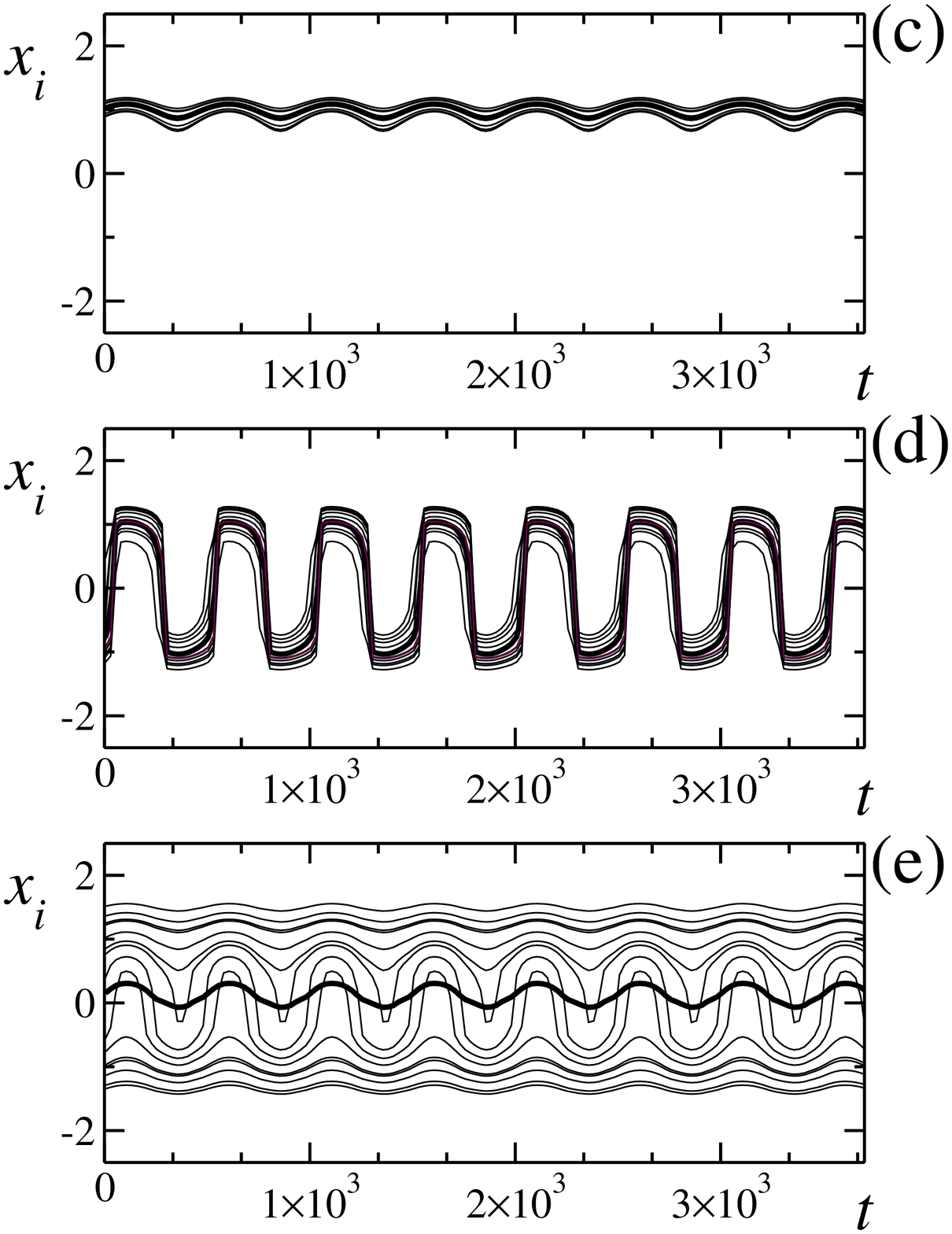}
\end{center}
\caption{\label{figure1}Diversity induced resonance in the $\phi^4$ model. In panel (a) we show the magnetization $m$ as a function of the diversity $\sigma$ for the $\phi^4$ model with quenched disorder defined in Eqs.(\ref{eq:phi4}). The parameters are $a=1$, $C=1$ and the values of $\eta_i$ are taken from a Gaussian distribution $g(\eta)$ of zero mean and variance $\sigma^2$. The line is the prediction of the mean-field theory and the symbols are the results of numerical simulations of the dynamical equations (\ref{eq:phi4}) for system sizes $N=50,10^2,10^3,10^4$ (the transition sharpens as $N$ increases). Panel (b) plots the spectral amplification factor $\eta$ as a function of diversity when the same system is forced by a periodic forcing of  amplitude $A=0.2$ and period $\Omega=500$. The line is the prediction of an adiabatic theory\cite{TMTG:2006} (not explained in the main text) and the symbols correspond to the numerical simulations in the case $N=10^3$. In the right panels, we show some representative trajectories $x_i(t)$ (thin lines) and the average trajectory $m(t)$ (thick line) in the case of $\sigma=0.20$ (panel c), $\sigma=0.54$ (panel d) and $\sigma=2.0$ (panel e); the other parameters as in panel (b). Note the wide variations of $m(t)$ in panel (d), corresponding to the optimal response to the external forcing.}
\end{figure}

In this section we review the disordering effect that diversity has on the $\phi^4$ model defined above. We use the all-to-all coupling where a full analytical understanding is possible. The all-to-all coupling assumption simplifies the problem and allows one to reduce it to a one variable. This is basically the Weiss mean-field treatment which is exact in the case of global coupling. Let us introduce the global variable $m(t)$ as:
\begin{equation}
m(t)=\frac{1}{N}\sum_{i=1}^Nx_i(t).
\end{equation}
The ``magnetization" is defined as the time average of this global variable:
\begin{equation}
m=\langle m(t)\rangle.
\end{equation}
The coupling between the $x_i$ variables appears only through this collective variable:
\begin{equation}
\frac{d x_i}{dt}=(a-C)x_i-x_i^3+Cm+\eta_i.
\end{equation}
This can be interpreted as a relaxational dynamics $\frac{d x_i}{dt}=-\frac{\partial V(x_i)}{\partial x_i}$ with a potential:
\begin{equation}
V(x_i)=\frac{C-a}{2}x_i^2+\frac{1}{4}x_i^4-(Cm+\eta_i)x_i.
\end{equation}
In the limit $t\to\infty$ the variable $x_i$ will tend to one of the minima of $V(x_i)$. We restrict ourselves from now on to the case $C\ge a$ for which the potential $V(x_i)$ has a single minimum, hence avoiding the possible metastable states that could occur otherwise. For fixed values of $m$ and $\eta_i$ the variable $x_i$ will tend to the unique solution of the cubic equation:
\begin{equation}
x_i^3+(C-a)x_i=Cm+\eta_i.
\end{equation}
The explicit solution is $x_i(\eta_i,m)=-\frac{\gamma}{u_i}+u_i$ with the notation $u_i=\left(\alpha_i+\sqrt{\gamma^3+\alpha_i^2}\right)^{1/3}$ and $\gamma=(C-a)/3$, $\alpha_i=(Cm+\eta_i)/2$. To determine the value of the mean-field variable $m$ we use the self-consistency relation (the subindex $i$ is now dropped from the notation):
\begin{equation}
m=\langle x \rangle=\int d\eta\,g(\eta)\,x(\eta,m),
\end{equation}
where, assuming self-averaging, we have replaced the sum over variables by an average over the realizations of the diversity variables $\eta$. Using the symmetry property of the distribution $g(\eta)$ the self-consistency relation can be expanded for small $m$:
\begin{equation}\label{eq:self_const}
m=F_1m+F_3m^3,
\end{equation}
where $F_1$, $F_3>0$ are coefficients that depend of $a$, $C$ and $\sigma$. As in the standard Weiss theory, this equation can have one or three solutions depending on the value of  $F_1$. If $F_1>1$ then the only solution corresponds to the disordered phase $m=0$. For $F_1<1$ there are two additional solutions $m=\pm m_0$, which correspond to the ordered phase. A detailed analysis shows that for fixed $a$ and $C$ the ordered solution $m\ne 0$ appears for a diversity $\sigma$ smaller than a critical value $\sigma_c$, while  a diversity $\sigma>\sigma_c$ only admits the disordered solution $m=0$. In this sense, diversity has a similar role to noise since a large diversity destroys the ordered state. The phase diagram is plotted in figure \ref{figure1} panel (a), in the case of a Gaussian distribution for $g(\eta)$ and $a=1$, $C=1$. In this case, the critical point can be computed as $\sigma_c=\left[\frac{\Gamma(1/6)}{ 2^{1/3}3\pi^{1/2}}\right]^{3/2}=0.7573428\dots$.

\subsection{The constructive role of diversity}\label{phi4-ord}

We now consider the same model under the influence of a periodic forcing. In the case that the dynamics is driven by noise (instead of disorder) it is well known the existence of stochastic resonance, i.e. an optimal value of the noise intensity for which the response of the system reaches a maximum value\cite{GHJM:1998}. We now demonstrate the existence of a similar effect under the presence of quenched noise. In the globally coupled case, we study the response to a periodic forcing of weak amplitude $A$ and frequency $\Omega$:
\begin{equation} \label{eq:phi4signal}
\frac{d x_i}{dt}=ax_i-x_i^3+\frac{C}{N}\sum_{j=1}^N(x_j-x_i)+\eta_i + A \, \sin (\Omega t).
\end{equation}
By a {\sl weak} amplitude $A$ we mean that its effect is such that the units make small oscillations around one of the equilibrium positions $x=\pm m_0$.
The global response is quantified by the average position of the units $m(t)=\frac{1}{N}\sum_i x_i(t)$ for which we now derive an evolution equation. By averaging the previous equation over the whole population, we obtain in the limit of large $N$:
\begin{equation}
\frac{dm}{dt}=am-\frac{1}{N}\sum_ix_i^3+A\, \sin (\Omega t).
\end{equation}
We now introduce $\delta_i=x_i-m$ and assume that the values of $\delta_i$ are evenly distributed around $0$. This leads to:
\begin{equation}
\frac{dm}{dt}=m(a-3\Delta)-m^3+A\, \sin (\Omega t),
\end{equation}
with $\Delta=\frac{1}{N}\sum_i\delta_i^2$. Hence, in the absence of forcing, $m(t)$ follows a relaxational dynamics in an effective potential:
\begin{equation} \label{eq:Vm}
V(m)=\frac{3\Delta-a}{2}m^2+\frac{1}{4}m^4.
\end{equation}
The potential barrier between the two minima is $\frac{(a-3\Delta)^2}{4}$. If the units are all identical, $\Delta=0$ and the barrier takes its maximum value. Any source of diversity will lead to $\Delta>0$ and the height of this barrier {\sl decreases for increasing diversity} until it eventually disappears altogether if $\Delta=a/3$. Therefore, it is possible that a forcing of weak amplitude $A$ that is subthreshold in the absence of diversity, becomes suprathreshold when diversity increases. It is predicted then that the global response $m(t)$ will increase suddenly when the diversity reaches a given value. For a large diversity, though, the potential $V(m)$ becomes monostable and the system again executes oscillations around a single minimum, now located at $m=0$. The existence of this resonance phenomenon can be clearly seen in figure \ref{figure1}(b) where we plot the spectral amplification factor $\eta=4A^{-2}|\langle{\rm e}^{i\Omega t}m(t)\rangle|^2$ as a function of the diversity $\sigma$.

In panels (c-e) of figure \ref{figure1} we plot some representative trajectories for the individual units $x_i(t)$ as well as the mean $m(t)$. In the case of small diversity, it can be seen that the units execute small oscillations around the same minimum following the external forcing (panel c). As the diversity increases over a critical value, the amplitude suddenly increases (panel d). Finally, for too large diversity, each unit now executes small oscillations, but each one oscillates around a different location and, hence, the amplitude of the oscillations of the global variable $m(t)$ decreases (panel e), in agreement with the previous discussion.

\section{Common firing in the active rotator model}\label{actrot}

Dynamical systems on a circle have been a paradigm to study synchronization phenomena \cite{YK:1984}.  This is due to their simplified dynamics that, in some cases, allows for analytical results. Nevertheless their simplicity, the results in those systems turn out to be broad applicability. In this section, we will focus on a system called {\em active rotator}. It exhibits either an oscillatory or excitable behavior, depending on the system parameters. The model we will consider is a globally coupled set of active rotators \cite{YK:1975,SK:1986} whose dynamics is given by
\begin{equation}\label{eq:phidot}
\dot{\phi_i} = \omega+\eta_i - \sin \phi_i + \frac{C}{N} \sum_{j=1}^N .
\sin\left(\phi_j-\phi_i \right).
\end{equation}
The global coupling of strength $C$ is written such that it is $2\pi$-periodic. The parameter $\omega_i=\omega+\eta_i$ is the natural frequency of the $j-$th unit. For uncoupled units, if $\omega_i < 1$, the unit is excitable: there are two-fixed points (one stable and another unstable). Any perturbation such that the system overcomes the unstable fixed point produces a {\em firing} of the unit: a large excursion until it arrives to the stable fixed point. On the other hand, if $\omega_i > 1$, there are no fixed points and the unit is in the oscillatory regime. The values of the natural frequencies are drawn from a probability distribution function $g(\eta_i)$, of mean $\langle\eta_i\rangle=0$ and correlations $\langle \eta_i\eta_j\rangle=\sigma^2\delta_{ij}$.  We consider the case $\omega<1$, such that when $\sigma=0$, all the systems are in the excitable regime and, in the absence of perturbations, they will all stay in perfect order at the stable equilibrium point. This order is degradated by the presence of diversity that makes each unit act differently from the others. We first study how the order decreases with diversity.

\subsection{The disordering role of  diversity}\label{actrot-dis}
Order in the position of the units, can be measured by means of the Kuramoto order parameter.  Let us define the (complex) variable $\rho(t){\rm e}^{i\Psi(t)}$ as the location of the center of mass of the rotators:
\begin{equation}\label{eq:kmoto}
\rho(t)e^{\imath \Psi(t)} = \frac{1}{N}\sum_{i=1}^N
e^{\imath \phi_i(t)}
\end{equation}
The Kuramoto order parameter, defined as the time average of this variable,
\begin{equation}\label{ordpar}
\rho {\rm e}^{\imath \Psi}= \left\langle\rho(t)e^{\imath \Psi(t)} \right\rangle,
\end{equation}
plays somehow the role of the magnetization $m$ in the $\phi^4$-model. The argument $\Psi$ is the global phase of the system. The modulus $\rho$ is a measure of the {\em order} in the position of the particles: it is equal to $1$ if all the units have the same phase, and it is $\rho=0$ for a situation in which the units are uniformly distributed around the circle. However, in the excitable regime there are two different dynamical regimes that could give rise to a value $\rho>0$:  a dynamical one, in which the units fire pulses synchronously, and  a static one, in which all the units are at rest in the stable fixed point.

\begin{figure}
\begin{center}
\includegraphics[width=7cm]{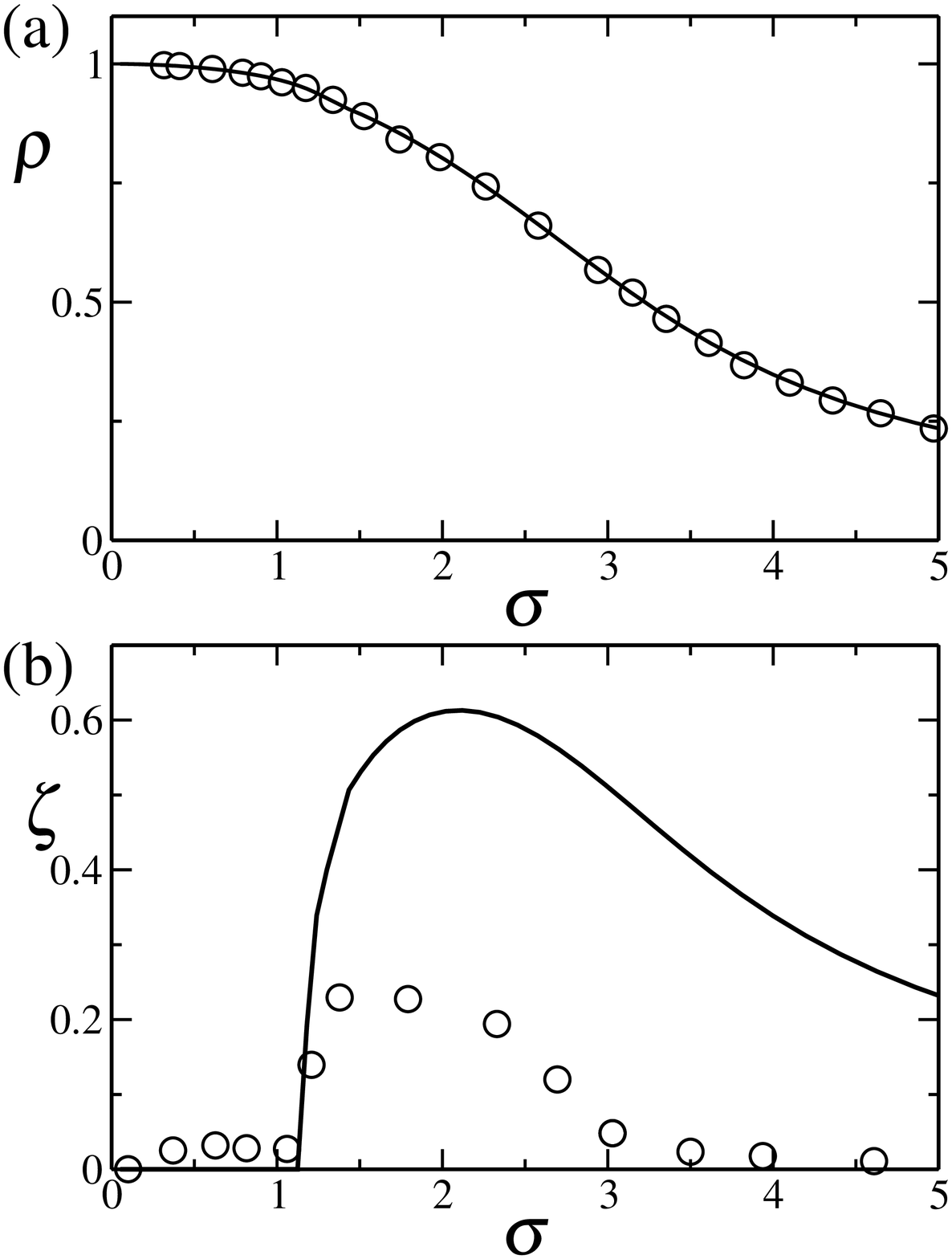}
\includegraphics[width=7cm]{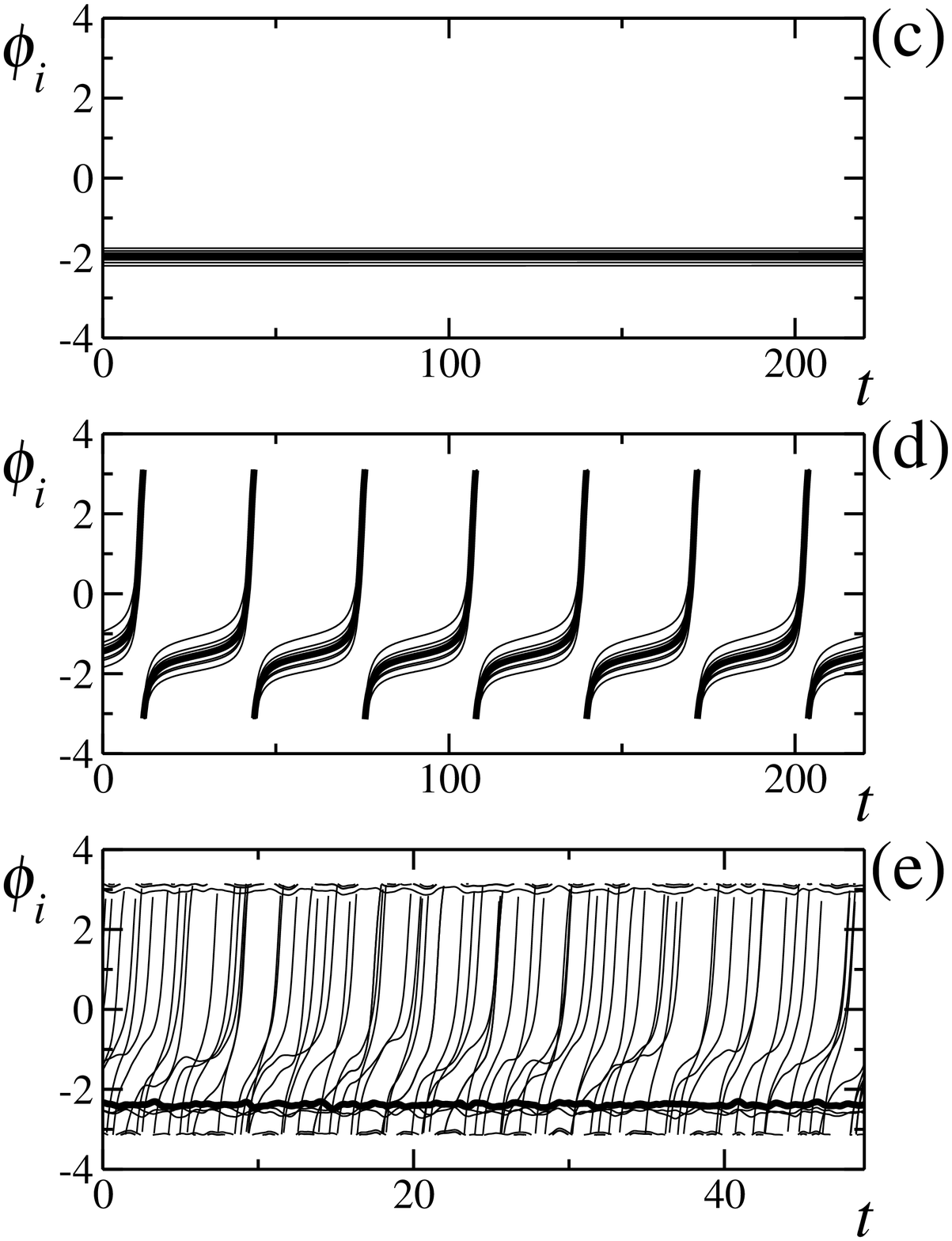}
\end{center}
\caption{\label{figure2}Common firing in the active-rotator model.  In
panel (a) we show the Kuramoto order parameter $\rho$ as a function of
the diversity $\sigma$ for the active-rotator model with quenched
disorder defined in Eqs.\ref{eq:phidot}. The parameters are
$\omega=0.95$, $C=4$, $N=400$ and the values of $\eta_i$ are taken from
a Gaussian distribution of zero mean and variance $\sigma^2$. The line is the mean-field prediction obtained by solving the self-consistent equation (\ref{scon}) and the dots are the results of numerical simulations of the dynamical equations (\ref{eq:phidot}). Panel (b) plots the Shinomoto-Kuramoto order parameter $\zeta$ as a function of diversity. A transition to a state in which the units pulse synchronously can be observed by the non-zero value of $\zeta$ starting around $\sigma=1.164$. The line is the theoretical prediction Eq.(\ref{eq:zeta}) and the dots come from the numerical simulations. In the right panels, we show some representative trajectories $\phi_i(t)$ (thin lines) and the
average phase $\Psi(t)$ (thick line) in the case of $\sigma=0.63$ (panel
c), $\sigma=1.73$ (panel d) and $\sigma=3$ (panel e). Note the
transition from the quiescent state (c) to a situation of common pulsing
(d) and then to incoherent pulsing (e).}
\end{figure}

We now show how to compute $\rho$. A straightforward algebra leads to dynamical equations for the angles $\phi_i$ in which  the coupling between units appears only through the global variables, $\rho$, $\Psi$, as:
\begin{equation}
\dot{\phi_i} = \omega_i - R\sin (\phi_i-\alpha),
\end{equation}
where $R=(1+2C\rho\cos\Psi+c^2\rho^2)^{1/2}$ and $\tan\alpha=\frac{c\rho\sin\Psi}{1+c\rho\cos\Psi}$. We now make the approximation of constant values for $\rho$ and $\Psi$. According to this equation, the rotators split in two categories: (i) those for who the natural frequency satisfies $|\omega_i|<R$ are in the excitable regime and  the probability density function of the angle distribution is a delta function centered around the stable angle $f(\phi_i)=\delta(\phi_i-\phi_i*)$ with $\phi_i^*=\arcsin(\omega_i)/R$; (ii)  those for which $|\omega_i|>R$ are in the oscillatory regime and the probability distribution is inversely proportional to the angular velocity $f(\phi_i)\propto |\dot \phi_i|^{-1}$. Computing the normalization constant we have:
\begin{equation}
f(\phi_i)=\cases{
\delta(\phi-\phi_i^*) & $|\omega_i|<R$ ,\cr
\frac{\sqrt{\omega_i^2-R^2}}{2\pi}\frac{1}{\omega_i-R\sin(\phi_i-\alpha)}&  $\omega_i> R$,\cr
\frac{\sqrt{\omega_i^2-R^2}}{2\pi}\frac{1}{-\omega_i+R\sin(\phi_i-\alpha)} & $\omega_i< -R$.
}
\end{equation}
This, in turn, allows to find the average value that appears in the definition of the Kuramoto order parameter as $\langle {\rm e}^{i\phi_i}\rangle=F(\eta_i,\rho,\Psi)$, with 
\begin{equation}
F(\eta_i,\rho,\Psi)={\rm e}^{i\alpha}\times\cases{
\sqrt{1-\frac{\omega_i^2}{R^2}}+\imath\frac{\omega_i}{R} & $|\omega_i|<R$, \cr
\imath\left(\frac{\omega_i}{R}-\sqrt{\frac{\omega_i^2}{R^2}-1}\right) &$\omega_i> R$,\cr
\imath\left(\frac{\omega_i}{R}+\sqrt{\frac{\omega_i^2}{R^2}-1}\right)  & $\omega_i< -R$.
}
\end{equation}
Finally, the order parameter is found by solving the consistency equation (the subindex  $i$ is now dropped from the notation):
\begin{equation}\label{scon}
\rho {\rm e}^{\imath\Psi}= \left\langle e^{\imath \phi} \right\rangle=\int d\eta \,g(\eta)F(\eta,\rho,\Psi).
\end{equation}
This equation has to be solved numerically. In figure \ref{figure2}(a) we plot $\rho$ versus the diversity $\sigma$ in the case of a Gaussian distribution for $g(\eta)$ and the values $\omega=0.95$, $C=1$ together with the results of numerical simulations of the dynamical equations \ref{eq:phidot}. It can be seen that the order parameter decreases monotonically as the diversity increases, although there is no sharp phase transition to a state of $\rho=0$. This is the usual role of disorder.  In the next subsection, we will show how diversity can induce a common firing of the active rotators.

\subsection{The constructive role of  diversity}\label{actrot-ord}
In order to analyze the global behavior we  derive evolution equations for the average variables $\rho(t)$ and $\Psi(t)$. We start by taking the time derivative of Eq.(\ref{eq:kmoto}) and introducing the deviation with respect to the average angle behavior as $\delta_i(t)=\phi_i(t)-\Psi(t)$. By expanding ${\rm e}^{i\delta_i}=1+i\delta_i+\mathcal{O}(\delta_i^2)$, we are led to $\dot \rho(t)=\mathcal{O}(\delta_i^2)$ and\cite{TSTC:2006}:
\begin{equation}
\dot\Psi=\frac{\omega}{\rho}-\sin\Psi+\mathcal{O}(\delta_i^2).
\end{equation}
This remarkable equation shows that the global phase $\Psi$ follows the same dynamics than a single rotator but with an effective natural frequency $\omega/\rho$. In the case of no diversity, $\sigma=0$, and $\omega<1$, all the rotators are in the rest state and $\rho=1$. As soon as some disorder is introduced in the rotators, the order parameter $\rho$ decreases such that when $\rho=\omega$ the global phase $\Psi$ becomes oscillatory. This is a true phase transition between a static phase and a dynamic phase in which the rotators pulse synchronously. As the value of $\rho$ changes continuously at the transition point it can not be used to characterize the phases. To distinguish between the two phases, Shinomoto and Kuramoto introduced another order parameter, $\zeta$, in which the mean value of the center of mass is subtracted before taking the modulus\cite{SK:1986}:
\begin{equation}
\zeta=\left\langle\left|\rho(t){\rm e}^{\imath\Psi(t)}-\left\langle\rho(t){\rm e}^{\imath\Psi(t)}\right\rangle\right|\right\rangle.
\end{equation}
If we approximate $\rho(t)$ by a constant value, it is $\zeta\approx\rho\left\langle\left|{\rm e}^{\imath\Psi(t)}-\left\langle{\rm e}^{\imath\Psi(t)}\right\rangle\right|\right\rangle$, the average over time can be performed as an average over the distribution $f(\Psi)$ of angles $\Psi$. As in the case of a single rotator the distribution for $\omega<\rho$ is a delta-function centered around the stable equilibrium value $\Psi^*=\arcsin(\omega/\rho)$, while for $\omega>\rho$ is is inversely proportional to the angular speed, $f(\Psi)\propto |\dot \Psi|^{-1}$. This allows to perform the angular average with the final result:
\begin{equation}\label{eq:zeta}
\zeta=\cases{
0 & $\omega\le\rho$,\cr
\frac{2}{\pi}\sqrt{2(\omega-\sqrt{\omega^2-\rho^2})(\omega+\rho)}K\left(\frac{2\rho}{\rho-\omega}\right) & $\omega>\rho,$
}
\end{equation}
where $K(x)$ is the complete elliptic integral of the first kind.
We plot in figure \ref{figure2}(b) the order parameter $\zeta$ as a function of the diversity $\sigma$. Now it is clear the transition from the static state to a regime of common pulsing at a critical value $\sigma_c\approx1.164$.  In the right panels of figure \ref{figure2} we plot some representative trajectories for the individual units $\phi_i(t)$ as well as the mean angle $\Psi(t)$. For low diversity, the units are at rest in their stable points (panel c). As the diversity $\sigma$ crosses the critical value $\sigma_c$ we can observe the synchronized firings (panel d). Finally, as the diversity increases further, the units start to fire in an unsynchronized manner (panel e).

\section{Conclusions}\label{conclusions}

In this paper we have reported two effects in which some level of  diversity induces a collective effect in dynamical systems. In the case of diversity induced resonance, the global response, as measured by the spectral amplification factor $\eta$ of the average variable $m(t)$ shows a clear maximum as a function of the diversity $\sigma$. The mechanism of the resonance is particularly simple: in the homogeneous case  the forcing is subthreshold for all units; as the diversity increases a fraction of the units are able to respond to the external forcing during half a period and another fraction of the units respond
during the next half period; through the coupling terms, the sensitive units are able to pull the rest of the units in the direction of the forcing; for too large diversity, the favorable units can not overcome the effect of the adverse ones.

In the active rotators system, we have shown the existence of a transition from the quiescent state where all units remain in the fixed stable point, to a regime of common pulsing induced by an increased in the diversity. Again, the mechanism is very simple: as the diversity increases, a fraction of the units have natural frequencies in the oscillatory regime; those units, through the coupling terms, pull the remaining ones into the oscillatory behavior.

Those two effects show a constructive role of diversity and are the analog of the same effects induced by noise.  It is to be stressed that it is the lack of perfect order that induces the collective effect. Our studies show that it is actually the disorder in the units that produce those collective effects and that it does not really matter whether the disorder is induced by noise, diversity or some other source, such as a random component in the network of connectivities or other sources. Since both phenomena require only generic ingredients, we expect that they should be present in the dynamics of suitable physical and biological systems and we hope that this work fosters experimental research in this direction.

\acknowledgements

This paper is the result of direct work of the authors with P. Colet, J.D. Gunton, C.R. Mirasso and A. Scir\`e. We acknowledge financial support by the Ministerio de Educaci\'on y Ciencia  (Spain), FEDER projects FIS2004-5073, FIS2004-953 and the EU NoE BioSim (LSHB-CT-2004-005137).


\end{document}